\documentclass[conference]{IEEEtran}
\usepackage[utf8]{inputenc}

\usepackage[cmex10]{amsmath}		
\usepackage{amssymb}				
\usepackage[binary-units]{siunitx}	
\usepackage{enumitem}

\usepackage[capitalise]{cleveref}	
\usepackage[backend=biber, sorting=none, bibstyle=ieee,citestyle=numeric-comp]{biblatex}
\bibliography{references.bib}
\usepackage{url}		

\usepackage{graphicx}	

\ifCLASSOPTIONcompsoc	\usepackage[caption=false,font=normalsize,labelfont=sf,textfont=sf]{subfig}
\else					\usepackage[caption=false,font=footnotesize]{subfig}
\fi

\usepackage{bm}

\usepackage[table,xcdraw]{xcolor}	
\usepackage{multirow}				
\usepackage{pbox}					
\usepackage[]{algorithm, algpseudocode}

\makeatother

\usepackage[]{algorithm, algpseudocode}
\makeatletter
\renewcommand{\ALG@beginalgorithmic}{\small}

\newcommand{\settest}[1]{\renewcommand{\max}{maximize}}
\graphicspath{{Figures/}}
\usepackage{microtype} 
\usepackage[font=footnotesize]{caption}
\begin{document}

\title{Placement and Allocation of Communications Resources in Slicing-aware Flying Networks}

\author{
	\IEEEauthorblockN{André Coelho, Helder Fontes, Rui Campos, Manuel Ricardo}
	\IEEEauthorblockA{INESC TEC and Faculdade de Engenharia, Universidade do Porto, Portugal\\
		\{andre.f.coelho, helder.m.fontes, rui.l.campos, manuel.ricardo\}@inesctec.pt}
		}

\maketitle

\begin{abstract}
Network slicing emerged in 5G networks as a key component to enable the use of multiple services with different performance requirements on top of a shared physical network infrastructure. A major challenge lies on ensuring wireless coverage and enough communications resources to meet the target Quality of Service (QoS) levels demanded by these services, including throughput and delay guarantees. The challenge is exacerbated in temporary events, such as disaster management scenarios and outdoor festivities, where the existing wireless infrastructures may collapse, fail to provide sufficient wireless coverage, or lack the required communications resources. Flying networks, composed of Unmanned Aerial Vehicles (UAVs), emerged as a solution to provide on-demand wireless coverage and communications resources anywhere, anytime. However, existing solutions mostly rely on best-effort networks. The main contribution of this paper is SLICER, an algorithm enabling the placement and allocation of communications resources in slicing-aware flying networks. The evaluation carried out by means of ns-3 simulations shows SLICER can meet the targeted QoS levels, while using the minimum amount of communications resources.
\end{abstract}


\section{Introduction} \label{sec:Introduction}
\begin{figure}
	\centering
	\includegraphics[width=1\columnwidth]{"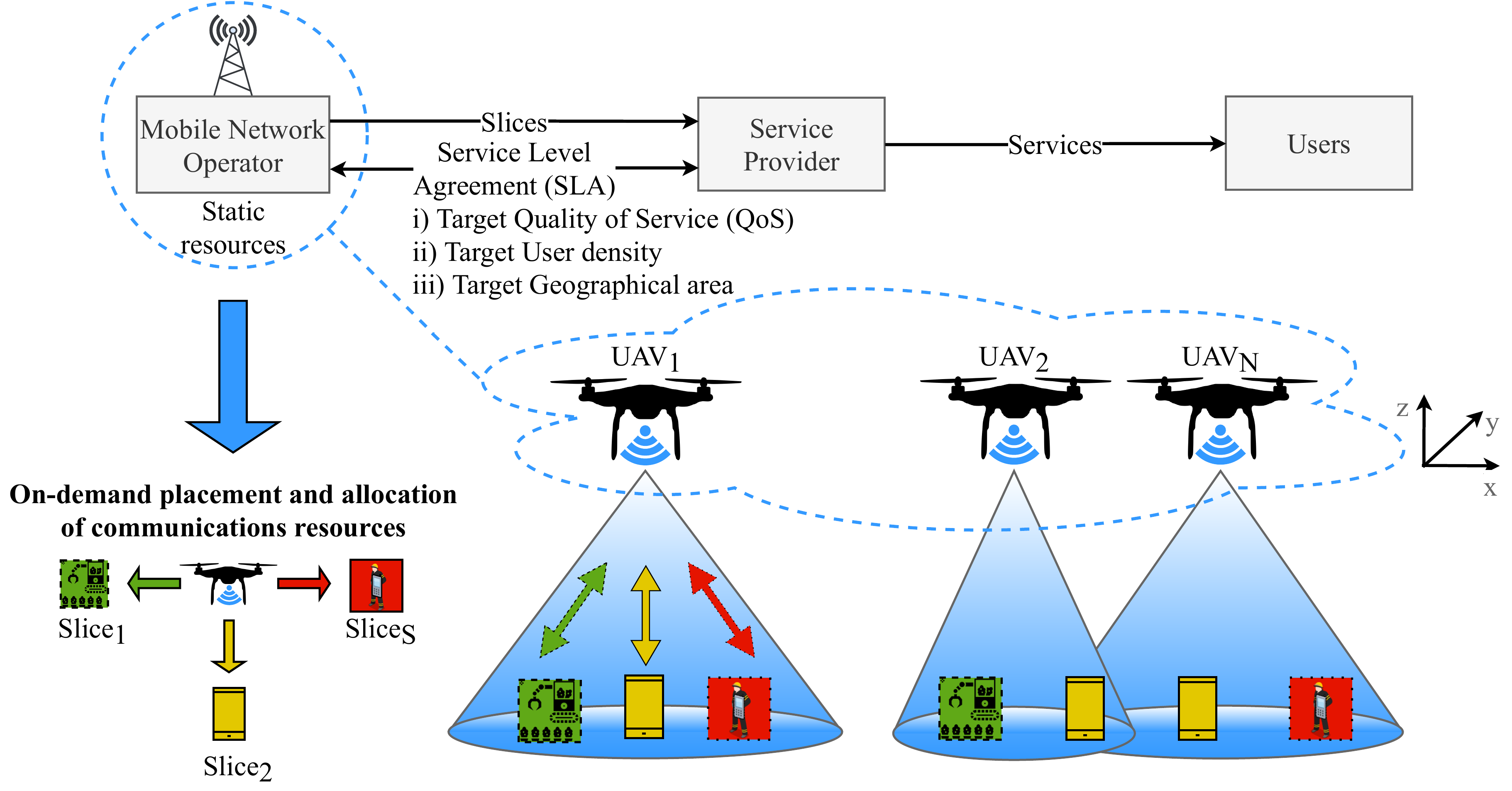"}
	\caption{Flying access network, composed of Unmanned Aerial Vehicles (UAVs), used by a Mobile Network Operator (MNO) to place and provide communications resources on-demand, so that $\left | S \right |$ network slices are made available to a Service Provider at target geographical areas.}
	\label{fig:sliced-flying-network}
\end{figure}

Network slicing aims at enabling the use of multiple services with different performance requirements on top of a shared physical network infrastructure~\cite{Luu2020Radio,Luu2020Coverage}. According to the 3\textsuperscript{rd} Generation Partnership Project (3GPP), a network slice is a logical network that provides specific network capabilities and target performance requirements, for instance, in terms of throughput and delay~\cite{3gppSlicing}. An important aspect related to the network slicing concept is its end-to-end nature, where a network slice is extended from the access network to the core network~\cite{Aleixendri2019}. Two major network entities may be associated to slicing: Mobile Network Operators (MNOs) and Service Providers or virtual MNOs. MNOs ensure the availability of the required communications resources, including the wireless and wired network infrastructure, such as cell sites, fronthaul and backhaul networks, and cloud data centers~\cite{Luu2020Coverage}. These resources can be owned and managed by the MNOs or leased by them from third-party infrastructure providers. Service Providers or virtual MNOs exploit and manage the network slices supplied by the MNOs, in order to meet the strict performance requirements of the services they offer, such as high definition video streaming, augmented and virtual reality, or smart metering. Service Providers and virtual MNOs act as tenants of the network infrastructure and provide services to their clients, performing the role of network users, as depicted in \cref{fig:sliced-flying-network}.\looseness=-1

In order to offer the Quality of Experience (QoE) expected by the users, the tenants request network slices to MNOs in the form of a Service Level Agreement (SLA). The SLA describes the service requirements at a high level, including Quality of Service (QoS) metrics, user density, and the geographical area where the services should be made available~\cite{Luu2020Coverage}.\looseness=-1

The literature on network slicing has been mainly focused on the resource allocation challenges, in order to ensure fairness and efficiency, while meeting the targeted QoS levels. Yet, a fixed network infrastructure is typically assumed~\cite{Luu2020Coverage}. A major challenge imposed to the MNOs lies on ensuring the availability of a dynamic wireless infrastructure able to provide wireless coverage and enough communications resources. This challenge is exacerbated in temporary events, such as disaster management scenarios and outdoor festivities, where the existing wireless infrastructures may collapse, fail to provide sufficient wireless coverage, or lack the required communications resources.\looseness=-1

In order to overcome this challenge, the use of flying networks, composed of Unmanned Aerial Vehicles (UAVs) acting as cellular Base Stations (BSs) and Wi-Fi Access Points (APs), emerged as a solution to provide agile, on-demand wireless coverage and communications resources, especially due to the UAVs' 3D positioning ability. Despite the ongoing research on UAV placement, existing solutions are focused on extending the wireless coverage and network capacity considering a best-effort approach, including our previous research~\cite{Coelho2019, Coelho2020, almeida2021joint}. Regarding network slicing in flying networks, state of the art works typically consider the UAVs acting as clients of telecom infrastructures and do not address the challenges regarding the placement and allocation of communications resources simultaneously~\cite{yuan2020airslice, Yang2020-A, Faraci2020, Cho2020}.\looseness=-1

The main contribution of this paper is SLICER, an algorithm enabling the on-demand placement and allocation of communications resources in slicing-aware flying networks composed of UAVs. SLICER allows to minimize the cost of deploying a  slicing-aware  flying  network by computing the minimum amount of communications resources, namely the number of UAVs acting as Flying Access Points (FAPs), needed to  meet the target QoS levels required by network slices made available in different geographical areas.\looseness=-1

The rest of this paper is organized as follows. 
\cref{sec:Related Work} presents the related work.
\cref{sec:System Model and Problem Formulation} describes the system model and formulates the problem.
\cref{sec:Slicer} details the placement and allocation of communications resources performed by the SLICER algorithm. 
\cref{sec:Performance Evaluation} refers to the performance evaluation of the flying network when SLICER is used.
Finally, \cref{sec:Conclusions} points out the main conclusions and future work.\looseness=-1

\section{Related Work \label{sec:Related Work}}
Network slicing emerged in 5G networks as a key component to ensure performance guarantees using the minimum amount of communications resources. In~\cite{kaloxylos2018survey}, a survey on network slicing for 5G networks is presented, including the 3GPP standardization process, whereas \cite{afolabi2018network} describes the key principles, enabling technologies, and open research challenges regarding network slicing. In the literature, several solutions have been proposed to meet the QoS levels of different network slices and ensure fairness and efficiency. In~\cite{vo2018slicing}, network slicing is formulated as an optimization problem, considering the backhaul network capacity, storage, and available bandwidth. However, only one BS in a fixed position is considered. In~\cite{navarro2020radio}, the resource allocation challenges in 5G+ access networks are discussed, including isolation, scalability, and efficiency. Due to the complexity in solving these problems, different approaches have been considered, including optimization~\cite{caballero2017multi}, game theory~\cite{d2018low}, evolutionary and heuristic algorithms~\cite{han2018slice, d2019slice}, and Machine Learning~\cite{xiang2020realization}. Still, these works are focused on maximizing the resource usage and minimizing costs, and do not address the placement of communications resources on-demand. A reference work that considers coverage constraints for providing network slices is presented in~\cite{Luu2020Coverage}. Yet, it assumes fixed BSs only. Overall, when it comes to wireless networks in general, the literature on network slicing has been focused on the resource allocation challenges, aiming at maximizing the resource utilization and minimizing the operation costs. However, a fixed network infrastructure is typically assumed. Providing wireless connectivity on-demand for ensuring coverage-aware network slices in dynamic environments has not been addressed so far~\cite{Luu2020Uncertainty}.\looseness=-1

Flying networks, composed of UAVs acting as Wi-Fi APs or cellular BSs, have emerged as a flexible and agile solution to provide on-demand wireless connectivity when there is no network infrastructure available or there is a need to enhance the coverage and capacity of existing networks~\cite{zeng2016wireless, chakraborty2018skyran, bor2016new}. Within this context, UAV placement algorithms have been proposed to determine the UAV positions that maximize the wireless coverage or the QoS and QoE offered to the users~\cite{cicek2019uav}. In particular, the solutions presented in~\cite{mozaffari2015drone, kalantari2017backhaul} aim at maximizing the area and number of users served, whereas the ones in~\cite{alzenad20173, panda2020efficient} focus on optimizing the QoS offered. The use of UAVs has also been envisioned under the Integrated Access and Backhaul (IAB) concept defined in 3GPP Release 16, which takes advantage of resources provided by BSs to establish backhaul networks~\cite{fouda2018uav}. Still, these solutions aim at improving the overall network performance by maximizing the Signal-to-Noise Ratio (SNR) and minimizing interference. In order to meet heterogeneous QoS levels, an on-demand density-aware 3D placement algorithm for a UAV acting as a BS is proposed in~\cite{lai2019demand}. It maximizes the number of users served, while promising guaranteed data rates, but considers only one UAV. In~\cite{donevski2021standalone}, a coverage-aware geometric placement approach for a UAV that provides connectivity to multiple users is proposed, considering the UAV altitude, cell size, and antenna’s beamwidth. Yet, this approach is targeted for areas that demand two network slices only, by placing a single UAV so that the average data rate of the users belonging to the enhanced Mobile Broadband (eMBB) slice is maximized, while ensuring coverage for the massive Machine-Type Communications (mMTC) slice. In~\cite{yang2021proactive}, an approach based on distributed learning and optimization is proposed, in order to provide an eMBB slice to the users, and an Ultra-Reliable and Low-Latency Communications (URLLC) slice for UAV control. A similar work is presented in~\cite{han2021dynamic}, in which multiplexing methods in time and frequency, and the effect of physical layer parameters, such as Modulation and Coding Scheme (MCS), are studied. Still, these works aim at enabling aggregate QoS guarantees and are not able to ensure different QoS levels for the same type of traffic being exchanged with different users.\looseness=-1  

When it comes to slicing in flying networks, the literature has been focused on UAVs acting as clients of telecom infrastructures~\cite{yuan2020airslice}. Existing works that take advantage of UAVs for providing on-demand network slices aim at improving energy-efficiency fair coverage~\cite{Yang2020-A, Faraci2020} and resource allocation efficiency~\cite{Cho2020}, but do not consider the placement and allocation of communications resources simultaneously. In~\cite{Skondras2020}, the UAV placement is optimized to improve the users' satisfaction when using different services, but no QoS guarantees are considered. In contrast, our work aims at paving the way to integrate UAVs into telecom infrastructures as on-demand communications resources that can be deployed anywhere, anytime, while meeting different QoS levels demanded by multiple network slices that coexist on top of a shared physical airborne infrastructure.\looseness=-1

\section{System Model and Problem Formulation~\label{sec:System Model and Problem Formulation}}
\begin{figure}
	\centering
	\setlength\abovecaptionskip{-0.01\baselineskip}
	\includegraphics[width=1\linewidth]{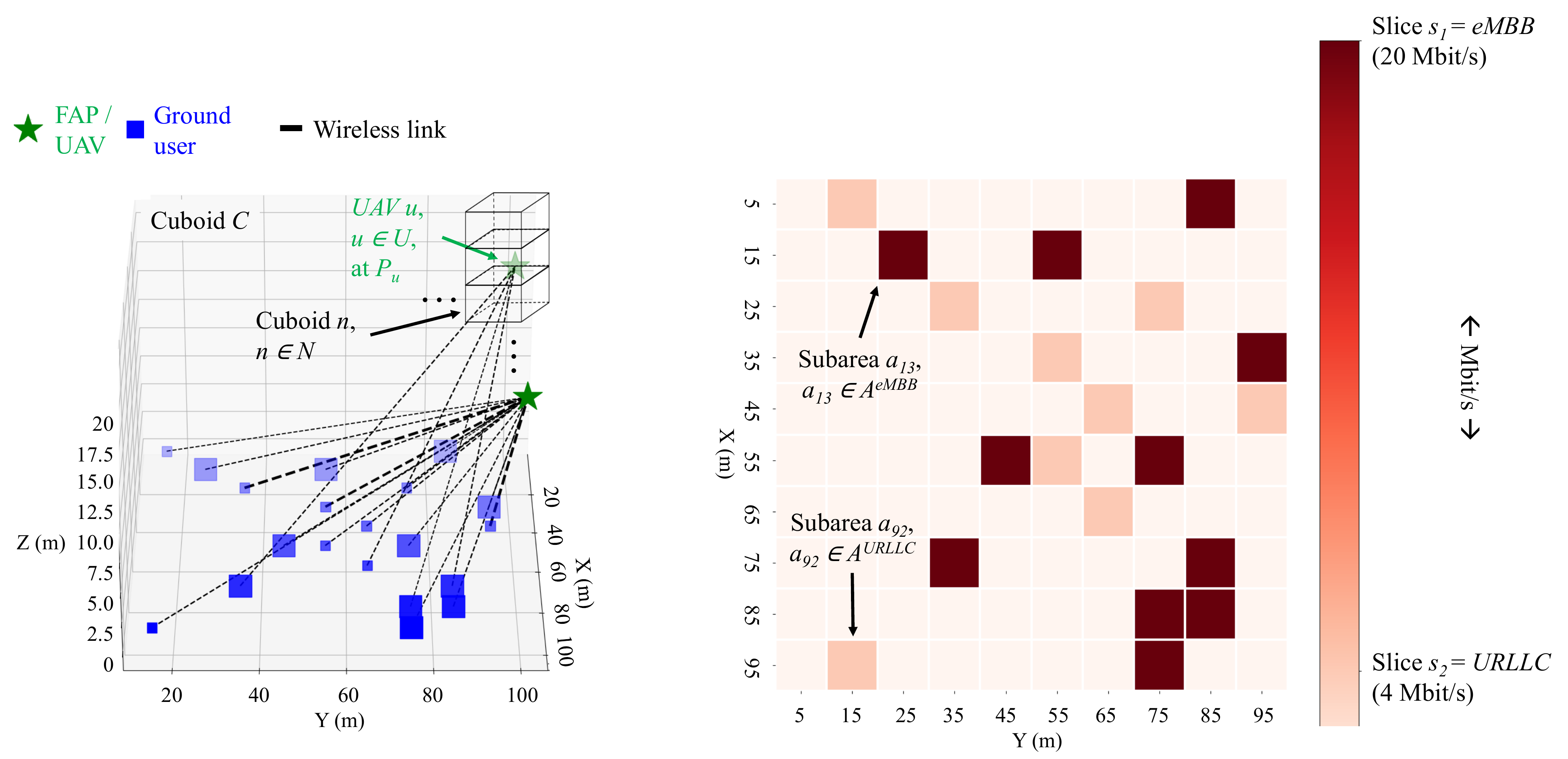}
	\caption{Illustrative networking scenario composed of multiple ground users (blue squares) served by two FAPs/UAVs $u \in U$ (green stars). The ground users are associated with two network slices $s \in S$ made available in different subareas $a \in A^s$ at the base of cuboid $C$ (2D representation on the right-hand side).}
	\label{fig:illustrative-networking-scenario}
\end{figure}

At time interval $t_k = k \cdot \Delta t, k \in \mathbb{N}_0$, where $\Delta t \gg \SI{1}{\second}$ is the flying network reconfiguration period, let $u \in U$ represent a UAV from the set of UAVs $U$ positioned inside a cuboid $C$ with dimensions $X$ long, $Y$ wide, and $Z$ high, as shown on the left-hand side of~\cref{fig:illustrative-networking-scenario}. Cuboid $C$ is divided into a set of $N$ equal and smaller fixed-size cuboids, where $n \in N$ represents a cuboid in which center a UAV may be located. $P_u=(x_u, y_u, z_u)$ is the position of UAV $u$. When used, UAV $u$ acts as a FAP that is in charge of providing wireless connectivity to the ground users placed in the base of cuboid $C$. The base of cuboid $C$ is divided into a set of fixed-size subareas (cf. right-hand side of~\cref{fig:illustrative-networking-scenario}). Let $a \in A$ represent a subarea, where $A$ is the set of subareas composing the base of cuboid $C$. Subarea $a \in A$ is centered at $P_a = (x_a, y_a, 0)$, where up to one and only one ground user is located.\looseness=-1

UAV $u$ has available a total number of $R_u$ wireless channels with constant bandwidth $B$, in \SI{}{\hertz}. We assume that $B$ is the minimum bandwidth per wireless channel, in \SI{}{\hertz}, that can be used to carry traffic to/from the ground users. This bandwidth depends on the wireless access technology employed. The management of the wireless link established between UAV $u$ and the Internet is beyond the scope of this paper; herein, we are focused on the wireless access network only.\looseness=-1

The wireless channel between UAV $u$ and the ground users is modeled by the Free-space Path Loss model. We assume the wireless channel is symmetric. The power $P_{R_{u,a}}$ received at $P_a$ from UAV $u$, in \SI{}{dBm}, is given by \cref{eq:free_space_model}, where $P_{T_u}$ is the transmission power of UAV $u$, in \SI{}{dBm}, and $G_{T_u}$ and $G_{R_a}$ are, respectively, the antenna gains of UAV $u$ and the ground user in subarea $a$, in \SI{}{dBi}. The path loss component $P_{L_{u,a}}$, in \SI{}{\decibel}, is computed by means of \cref{eq:path_loss}, where $d_{u,a}$ is the Euclidean distance between $P_u$ and $P_a$, in meters, $f_u$ is the carrier frequency used by UAV $u$, in \SI{}{\hertz}, and $c$ is the speed of light in vacuum, in \SI{}{\meter/\second}.\looseness=-1 

\begin{scriptsize}
\begin{equation}
    P_{R_{u,a}} = P_{T_u} + G_{T_u} + G_{R_a} - P_{L_{u,a}}
    \label{eq:free_space_model}
\end{equation}

\begin{equation}
    P_{L_{u,a}} = 20 \cdot log_{10}(d_{{u,a}}) + 20 \cdot log_{10}(f_u) + 20 \cdot log_{10}\left (\frac{4 \cdot \pi}{c}\right)
    \label{eq:path_loss}
\end{equation}
\end{scriptsize}

The SNR received at $P_a$ from UAV $u$, in \SI{}{\decibel}, is given by $SNR_{u,a} = P_{R_{u,a}} - P_{N_{u,a}}$, where $P_{N_{u,a}}$ is the noise power, in \SI{}{dBm}, which we assume to be constant for the channel bandwidth $B$. The capacity provided by each wireless channel is equal to the data rate associated with the Modulation and Coding Scheme ($MCS_{u,a}$) index used for the wireless link established between UAV $u$ and the ground user at $P_a$. The use of $MCS_{u,a}$ imposes a minimum $SNR_{u,a}$, considering a constant noise power $P_{N_{u,a}}$.\looseness=-1

The number of wireless channels provided by UAV $u$ to subarea $a \in A$ during time interval $t_k$ is denoted by $r_{u,a}(t_k)$. The number of wireless channels provided by UAV $u$ to all subareas $a \in A$ must be lower than or equal to the total number $R_u$ of wireless channels available at $UAV_u$, as defined in \cref{eq:number_rus}.\looseness=-1

\begin{scriptsize}
\begin{equation}
    \sum_{a \in A} r_{u,a}(t_k) \leq R_u, \forall u \in U
    \label{eq:number_rus}
\end{equation}
\end{scriptsize}

The number of UAVs serving subarea $a \in A$ during time interval $t_k$ is denoted by $K_a(t_k)$. We assume subarea $a \in A$ is served by one and only one UAV $u$, as stated in~\cref{eq:number_uavs_per_subarea}.\looseness=-1

\begin{scriptsize}
\begin{equation}
    \begin{aligned}
        K_a(t_k) = 1, \forall a \in A
    \end{aligned}
    \label{eq:number_uavs_per_subarea}
\end{equation}
\end{scriptsize}

The indicator function $1_u(t_k)$, defined in~\cref{eq:rus-to-slide}, denotes whether UAV $u$ serves any subarea $a\in A$ during time interval $t_k$.\looseness=-1 

\begin{scriptsize}
\begin{equation}
\begin{aligned}
1_u(t_k) =
\left\{ 
  \begin{array}{ c l }
    1,  &   \quad   \textrm{if } \sum_{a \in A} r_{u,a} (t_k) > 0, \forall u \in U\\
    0,  &   \quad   \textrm{otherwise}
  \end{array}
\right.
\end{aligned}
\label{eq:rus-to-slide}
\end{equation}
\end{scriptsize}

Let us consider that a Service Provider rents a set of network slices from an MNO, in order to offer online services to the ground users located in area $A$, as depicted in \cref{fig:illustrative-networking-scenario}. Let $s \in S$ represent a network slice, where $S$ is the set of network slices. We assume that each subarea $a\in A$ occupied by a ground user is associated with a single network slice $s$, but a network slice $s$ can cover multiple subareas $a \in A$. As such, network slice $s$ enables the use of a service made available in area $A^s \subset A$. The area $A^s$ associated with network slice $s$ is the union of a set of fixed-size subareas $a \in A$.\looseness=-1

The average data rate available in subarea $a \in A^s$ when using a given number of wireless channels must be higher than or equal to the average data rate $T^s$ demanded by network slice $s$, as denoted in \cref{eq:data_rate_rus}. $c_{u,a} (t_k)$ represents the bidirectional network capacity provided by a wireless channel with constant channel bandwidth $B$, in terms of the amount of bit/s carried between UAV $u$ and the ground user located in area $a\in A^s$.\looseness=-1 

\begin{scriptsize}
\begin{equation}
    \begin{aligned}
    \sum_{u \in U} c_{u,a}(t_k) \cdot r_{u,a} (t_k) \geq T^s, 
    & \forall a \in A^s, \forall s \in S
    \end{aligned}
    \label{eq:data_rate_rus}
\end{equation}
\end{scriptsize}

\begin{figure}
	\centering
	\subfloat[Capacity versus SNR for an eMBB network slice, considering the target BER equal to 10\textsuperscript{-5}.]{
	\includegraphics[width=0.49\linewidth]{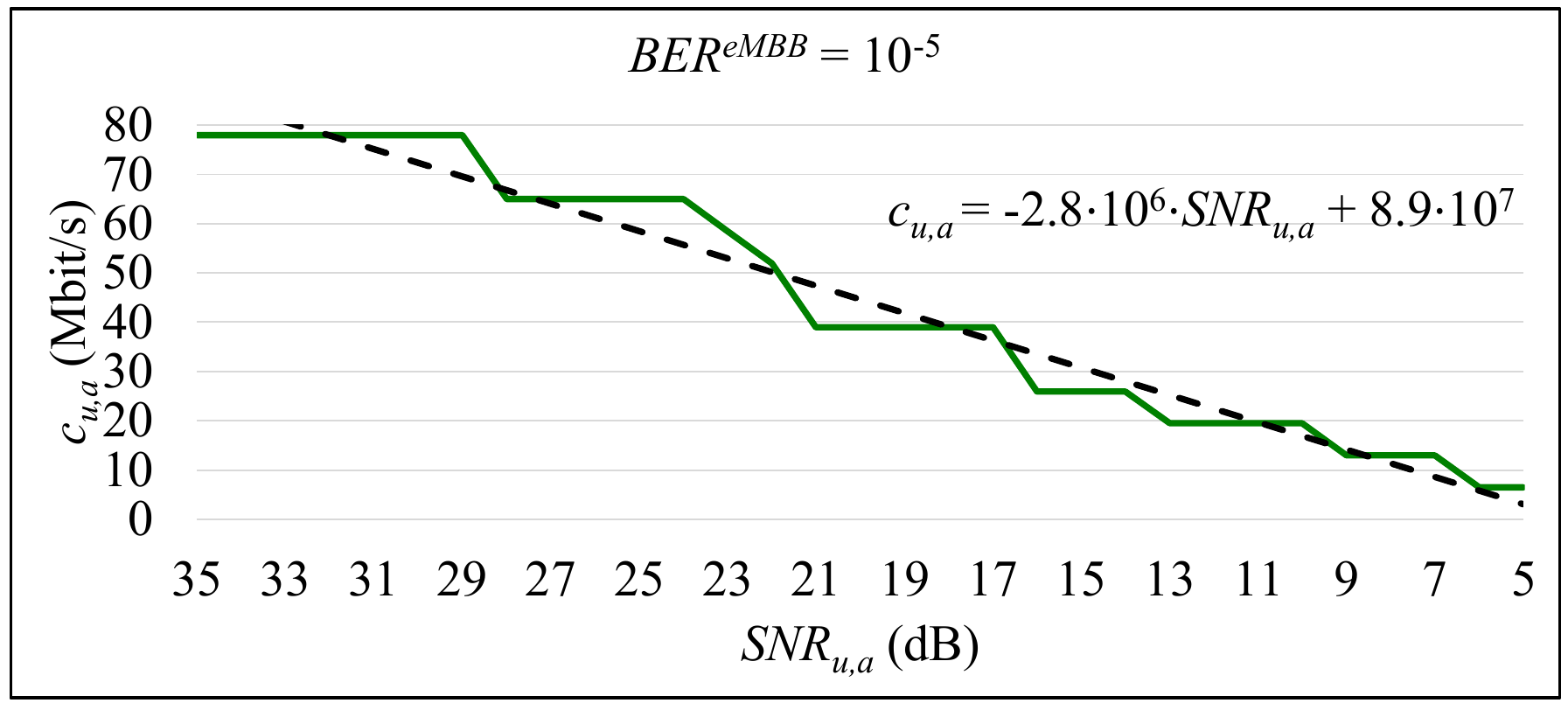}
	\label{fig:data-rate-snr-eMBB}}
	\hfill
	\subfloat[Capacity versus SNR for a URLLC network slice, considering the target BER equal to 10\textsuperscript{-10}.]{
		\includegraphics[width=0.425\linewidth]{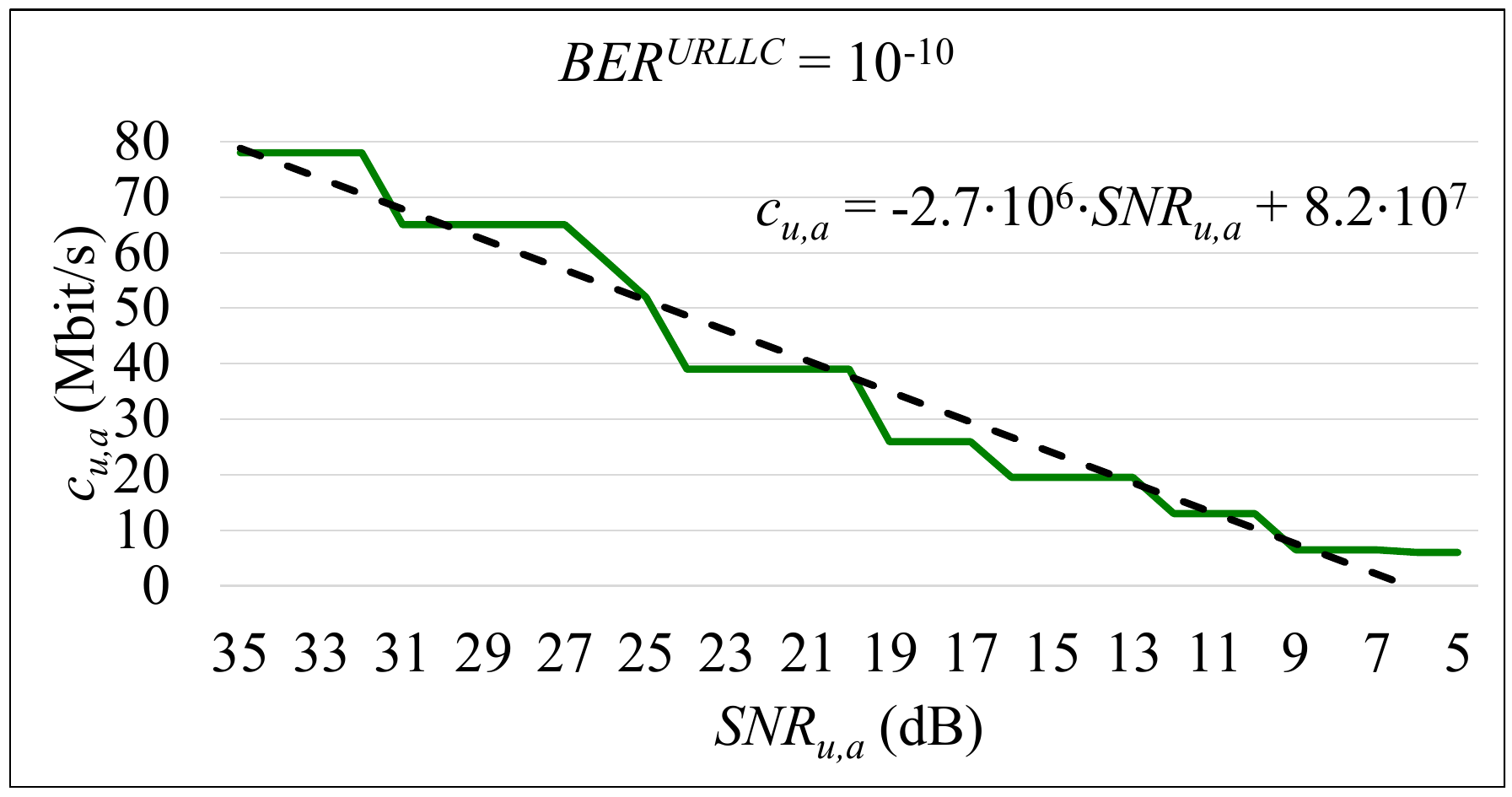}
		\label{fig:data-rate-snr-URLLC}}
	\hfill
    \caption{Wireless channel capacity modeled by linear regressions between the SNR and the data rate associated to the IEEE 802.11ac MCS indexes, considering \SI{20}{\mega\hertz} channel bandwidth.}
    \label{fig:data-rate-snr}
\end{figure}

The relation between the minimum $SNR_{u,a}$ from UAV $u$ in subarea $a \in A^s$ required for using $MCS_{u,a}$ is considered, taking into account target Bit Error Rate (BER) values according to the requirements of network slice $s$. For improved reliability, higher SNR values must be ensured, so that a lower BER is achieved \cite{Korrai2020}. For illustrative purposes, BER equal to 10\textsuperscript{-10} is considered for a URLLC network slice type, while BER equal to 10\textsuperscript{-5} is employed for an eMBB network slice type. The relation between $SNR_{u,a}$ and $MCS_{u,a}$ for different target BER values is presented in \cref{fig:data-rate-snr}, considering the IEEE 802.11ac standard, \SI{800}{\nano\second} Guard Interval (GI), and \SI{20}{\mega\hertz} channel bandwidth. Since the relation between $SNR_{u,a}$ and $MCS_{u,a}$ is step-wise (cf. solid green lines in~\cref{fig:data-rate-snr}), making the problem intractable and complex to solve mathematically, we model it as a continuous function using a linear regression (cf. dashed black lines in~\cref{fig:data-rate-snr}), which is a function that closely fits the data. $c_{u,a} (t_k)$ changes according to the location of subarea $a\in A^s$ and the position of UAV $u$, since both influence $SNR_{u,a}$.\looseness=-1

For the ground user in subarea $a \in A^s$, the traffic being forwarded by UAV $u$ is modeled by an M/D/1 queue $Q_{u,a}$ (Poisson arrival, deterministic service time, 1 server)~\cite{bershkas}. Traffic arrives at queue $Q_{u,a}$ with arrival rate $\lambda_{u,a}$ packet/s and is served with a service rate $\mu_{u,a}$ packet/s. The average delay $D_{u,a}(t_k)$ of a packet generated by the ground user in subarea $a\in A^s$ during time interval $t_k$ is computed using \cref{eq:avg_packet_delay}.\looseness=-1  

\begin{scriptsize}
\begin{equation}
    \begin{aligned}
        D_{u,a}(t_k) = \frac{1}{\mu_{u,a}} \cdot \frac{\rho_{u,a}}{2 \cdot \mu_{u,a} \cdot (1-\rho_{u,a})} \cdot 1_u(t_k), \\
        \forall u \in U, \forall a \in A^s, \forall s \in S
    \end{aligned}
    \label{eq:avg_packet_delay}
\end{equation}

where:
\begin{itemize}[label=]
    \item $\rho_{u,a} = \frac{\lambda_{u,a}}{\mu_{u,a}} < 1$, $\lambda_{u,a} \neq 0, \forall u \in U, \forall a \in A^s, \forall s \in S$
\end{itemize}
\end{scriptsize}

The average packet delay $D_{u,a}(t_k)$ must be lower than or equal to the maximum average packet delay $H^s$ associated with network slice $s$, as given by~\cref{eq:average_packet_delay}. 

\begin{scriptsize}
\begin{equation}
    \begin{aligned}
        D_{u,a} \cdot 1_u(t_k) \leq H^s, \forall u \in U, \forall a \in A^s, \forall s \in S
    \end{aligned}
    \label{eq:average_packet_delay}
\end{equation}
\end{scriptsize}

Herein, we consider average QoS values for illustrative purposes, but an SLA established with an MNO can also refer minimum values (e.g., lowest packet delay among all packet delay values) or median values (e.g., 50\textsuperscript{th} percentile of the frequency distribution of packet delays)~\cite{Janevski2017}.\looseness=-1

The problem consists in minimizing the cost of deploying a slicing-aware flying network able to meet the coverage and QoS levels of any network slice $s \in S$, including a minimum average data rate $T^s$ and a maximum average packet delay $H^s$. Solving this problem includes determining the minimum number of UAVs to use, their 3D positions, and the number of wireless channels they provide to the subareas associated with the network slices. We assume that there is a set of $N$ predefined 3D positions where potential UAVs are placed. Placing UAV $u \in U$ at $P_u$, during time interval $t_k$, has a fixed activation cost $F_u$, where $F_u$ is a constant associated with the usage of UAV $u$. This activation cost may be defined according to multiple criteria, such as the cost of the hardware carried on board and the operating cost of each UAV. The optimization problem, including the objective function, is defined in \cref{eq:optimization-problem}.\looseness=-1

\begin{scriptsize}
  \begin{subequations}
    \begin{alignat}{10}
        & \!\underset{r_{u,a}(t_k), 1_u(t_k)}{\textrm{minimize}} && \sum_{u \in U} F_u \cdot 1_u(t_k), \forall a \in A \label{eq:objective-function}\\
         & \textrm{subject to:} \notag \\
         & && \hspace{-5em} \sum_{a \in A} r_{u,a}(t_k) \leq R_u, \forall u \in U\\
         & && \hspace{-5em} \sum_{u \in U} c_{u,a}(t_k) \cdot r_{u,a} (t_k) \geq T^s, \forall a \in A^s, \forall s \in S\\
         & && \hspace{-5em} D_{u,a} \cdot 1_u(t_k) \leq H^s, \forall u \in U, \forall a \in A^s, \forall s \in S\\
         & && \hspace{-5em} K_a(t_k) = 1, \forall a \in A
    \end{alignat}
    \label{eq:optimization-problem}
  \end{subequations}
\end{scriptsize}

The main notation used to formulate the problem addressed by SLICER is presented in~\cref{tab:notation}.

\begin{scriptsize}
  \begin{table}[!ht]
    \centering
    \caption{Main notation used to formulate the problem addressed by SLICER.}
    \label{tab:notation}
    \begin{tabular}{lp{7cm}lp{\columnwidth}}
      \hline
      \textbf{Symbol} & \textbf{Definition} \\
      \hline
      $u, a, s, t_k$   & Symbols representing respectively the UAVs, subareas, network slices, and time intervals\\
      $A$ & Area forming the base of cuboid $C$\\
      $A^s$ &   Set of subareas $a\in A$ where the network slice $s$ is made available\\
      $C$   &   Cuboid within which the set of UAVs $U$ can be positioned\\
      $P_u, P_a$ & Position of UAV $u$ and center of subarea $a\in A$\\
      $c_{u,a}$    &   Bidirectional network capacity provided by a wireless channel made available by UAV $u$ to the ground user at $P_a$, in \SI{}{bit/s}\\
      $G_{R_a}$ &  Antenna gain of the ground user at $P_a$, in \SI{}{dBi}\\
      $G_{T_u}, P_{T_u}$ &  Antenna gain, in \SI{}{dBi}, and transmission power of UAV $u$, in \SI{}{dBm}\\
      $MCS_{u,a}$ & Modulation and Coding Scheme index used in the wireless link established between UAV $u$ and the ground user at $P_a$\\
      $P_{L_{u,a}}$ &   Path loss component between UAV $u$ and $P_a$, in \SI{}{\decibel}\\
      $P_{R_{u,a}}$ &   Power received at $P_a$ from UAV $u$, in \SI{}{dBm}\\
      $R_u$ &   Total number of wireless channels available at UAV $u$\\
      $r_{u,a}$ &   Number of wireless channels made available by UAV $u$ to subarea $a\in A$\\
      $SNR_{u,a}$ & SNR received at $P_a$ from $UAV_u$, in \SI{}{\decibel}\\
      \hline
    \end{tabular}
  \end{table}
\end{scriptsize}

\section{SLICER Algorithm \label{sec:Slicer}}
In this section, we detail SLICER, the proposed algorithm to solve the problem formulated in~\cref{sec:System Model and Problem Formulation}. SLICER generates a finite subspace of admissible solutions for the problem and uses a state of the art solver to determine the optimal solution. In addition, SLICER defines the communications resource allocation by means of a channel assignment approach that minimizes the overall bandwidth required.\looseness=-1

\begin{algorithm}
	\renewcommand{\thealgorithm}{A}
	\caption{-- SLICER algorithm}
	\begin{algorithmic}[1]
	\label{alg:slicer}
		\State Discretize cuboid $C$ into $N$ cuboids centered at $P_u$
		\State Compute $SNR_{u,a}$ for the wireless link available between each potential UAV $u$ and the ground user at $P_a$
		\State Compute the network capacity $c_{u,a}$ provided by a wireless channel with bandwidth $B$ to the ground user at $P_a$
		\State Solve the optimization problem defined in \cref{eq:optimization-problem} using a state of the art solver
		\State Assign the wireless channels that minimize the bandwidth used
		\State Reconfigure the flying network accordingly
	\end{algorithmic}
\end{algorithm}

Inspired by the capacitated facility location problem~\cite{Melkote2001}, a classical optimization problem that aims at selecting the best among potential locations for a factory or warehouse, we propose the SLICER algorithm, in order to place and allocate communications resources in slicing-aware flying networks.\looseness=-1 

Following~\cref{alg:slicer}, SLICER initially considers a set of $N$ smaller cuboids, each associated to a potential FAP deployed at $P_u$, and a given number of ground subareas, each occupied by a user in the central position. Each potential FAP can give rise to a real FAP, if it is part of the final solution. In turn, each ground user is characterized by a homogeneous known traffic demand and a maximum average packet delay, which are values associated to the SLA of a given network slice $s \in S$. Then, SLICER computes the SNR of the wireless link that can be established between each potential FAP and the ground user in the center of each subarea $a\in A^s$. After that, SLICER determines the network capacity achievable when using the minimum channel bandwidth that can be provided by each potential FAP. The minimum channel bandwidth is a configuration parameter that may be defined according to the communications technology used by the flying network -- e.g., \SI{20}{\mega\hertz} for IEEE 802.11 or in terms of the number of Orthogonal Frequency-Division Multiple Access (OFDMA) Resource Units (RUs) for IEEE 802.11ax or 5G New Radio. Assuming that each potential FAP has a given activation cost, as well as a limited number of wireless channels available, a state of the art solver is used by SLICER, in order to minimize the sum of the potential FAPs' activation costs, which includes identifying the potential FAPs/UAVs to actually use, and the number of wireless channels to be made available by each FAP. At the same time, the solver must ensure that the SLA associated with each network slice is met, while the capacity of each FAP in terms of the total number of wireless channels available is not exceeded. In its current version, SLICER uses the Gurobi\textsuperscript{TM} optimizer~\cite{gurobi} and considers 1000 € as the activation cost, defined based on a realistic cost for an off-the-shelf quadcopter UAV~\cite{droneshop2021}, as well as eight \SI{20}{\mega\hertz} wireless channels available per FAP, enabling up to \SI{160}{\mega\hertz} channel bandwidth. The resulting solution consists of the potential FAPs to use, which are associated to a known location, as presented in~\cref{fig:illustrative-networking-scenario}, and the minimum amount of wireless channels that they must provide to each subarea $a \in A$. When a potential FAP does not provide resources to any subarea $a\in A$, its activation cost is zero and the potential FAP does not give rise to a real FAP.\looseness=-1    

\begin{figure}
	\centering
	\subfloat[Resources required before channel assignment. Each row is associated to a subarea $a\in A^s$.]{
	\includegraphics[width=0.46\linewidth]{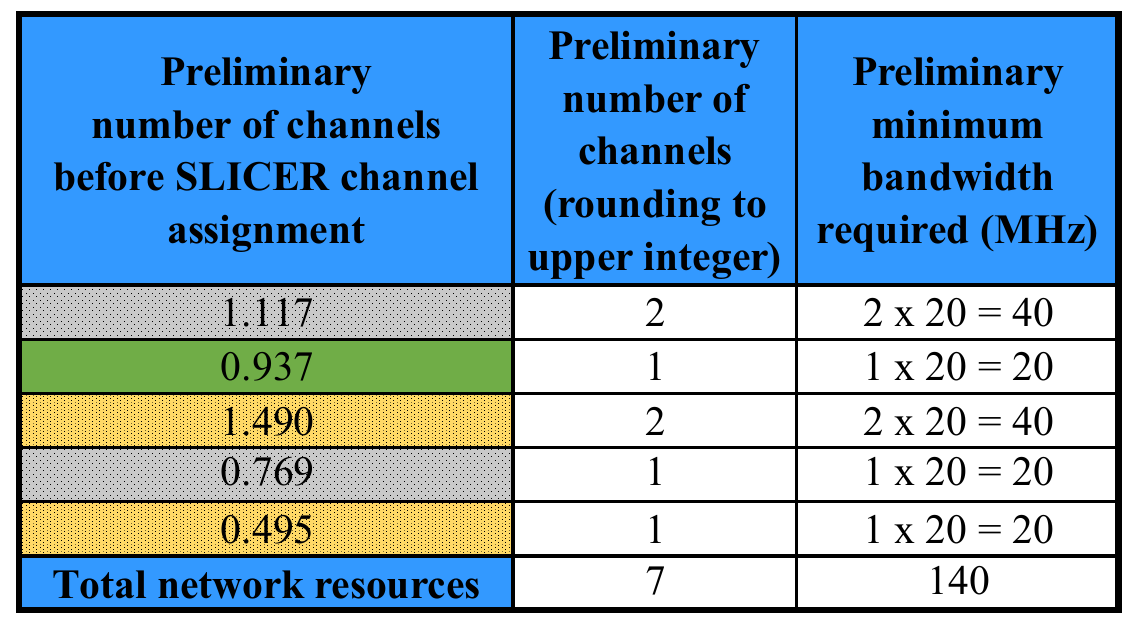}
	\label{fig:BW-allocation-before-channel-assignment}}
	\hfill
	\subfloat[Resources required after channel assignment. The quantities of wireless channels stated in~\cref{fig:BW-allocation-before-channel-assignment} are added together according to the color patterns.]{
		\includegraphics[width=0.46\linewidth]{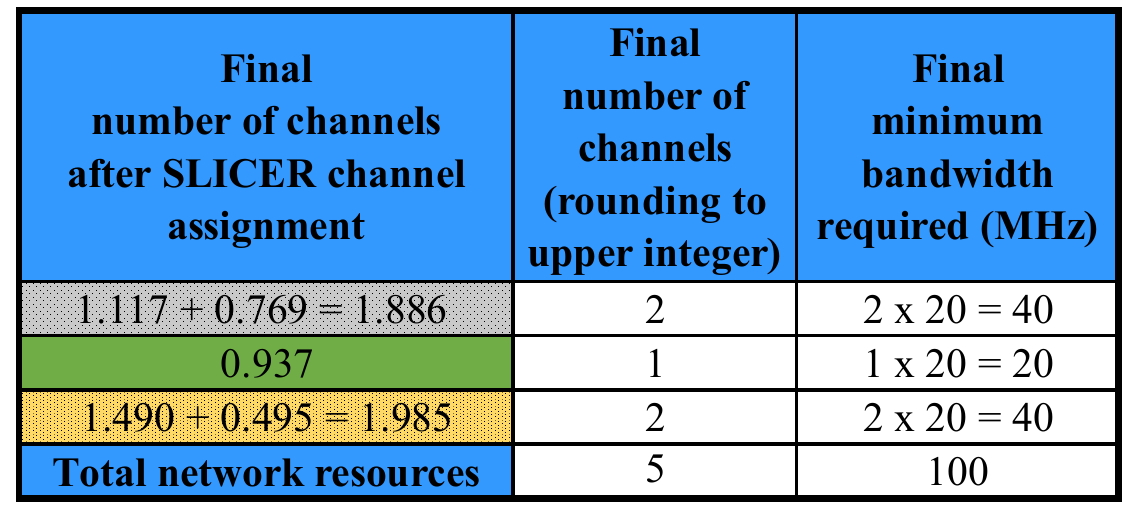}
		\label{fig:BW-allocation-after-channel-assignment}}
	\hfill
    \caption{Resource allocation performed by SLICER, considering the minimum channel bandwidth equal to \SI{20}{\mega\hertz}. 
    }
    \label{fig:BW-allocation-SLICER}
\end{figure}

Since a precise allocation of the network resources is not achievable in some wireless communications technologies, including IEEE 802.11, where the channel bandwidth must be an integer multiple of \SI{20}{\mega\hertz}, SLICER performs the resource allocation by means of a suitable channel assignment approach that minimizes the overall bandwidth used. For that purpose, different subareas are assigned with the same wireless channel, aiming at reducing the overall bandwidth wasted by sharing the spectral resources available. An illustrative example is presented in~\cref{fig:BW-allocation-SLICER} for a network slice $s$. Without the channel assignment approach considered by SLICER, the total bandwidth required is \SI{140}{\mega\hertz}, as presented in~\cref{fig:BW-allocation-before-channel-assignment}. Such baseline approach considers different \SI{20}{\mega\hertz} wireless channels assigned to each subarea $a\in A^s$. Please note that the total bandwidth required will increase as more subareas $a\in A^s$ are considered -- each row in~\cref{fig:BW-allocation-before-channel-assignment} corresponds to a subarea $a\in A^s$. However, taking into account that for each subarea $a\in A^s$ the bandwidth used is far from the full channel bandwidth available, this baseline channel assignment approach leads to a waste of spectral resources.\looseness=-1 

With the channel assignment performed by SLICER, the total bandwidth used is reduced to \SI{100}{\mega\hertz}. This is accomplished by using the maximum possible bandwidth of each wireless channel made available, while assigning the same wireless channel to the maximum number of subareas $a\in A^s$, as presented in~\cref{fig:BW-allocation-after-channel-assignment}. This channel assignment approach allows to overcome the limitations imposed by some communications technologies that make it difficult to accurately allocate the amount of communications resources required. In addition, it allows to reduce the spectral resources used, without compromising the QoS guarantees.\looseness=-1

\section{Performance Evaluation \label{sec:Performance Evaluation}}
The evaluation of the flying network performance when using SLICER is presented in this section. It refers to the simulation setup employed, the simulation scenarios considered, the performance metrics used, and the main simulation results obtained.\looseness=-1

\subsection{Simulation Setup}
In order to evaluate the performance achieved by the slicing-aware flying network when SLICER is employed, the ns-3 simulator~\cite{ns-3} was used. The ground users, performing the role of IEEE 802.11ac Stations (STAs), and the FAPs, acting as IEEE 802.11ac APs, were using a Network Interface Card (NIC) in Infrastructure mode. Up to $\left | S \right |$ wireless channels with up to \SI{160}{\mega\hertz} channel bandwidth, \SI{800}{\nano\second} GI, and a single spatial stream were employed for the wireless links. The \emph{IdealWifiManager} mechanism was in charge of automatically defining the data rate, since it allows to configure target BER values, according to the SNR of the wireless links, as considered in~\cref{sec:System Model and Problem Formulation}. For each networking scenario that was considered in the performance evaluation, SLICER was used to determine the minimum number of UAVs, their 3D positions, and the minimum bandwidth to allocate to each subarea $a\in A^s$.\looseness=-1

\subsection{Simulation Scenarios}\label{sec:Simulation Scenarios}
Three sets of five networking scenarios consisting of 5, 20, and 45 ground users, respectively, were considered. For each networking scenario, the ground users were randomly distributed among two different network slices (eMBB and URLLC) and randomly positioned on the base of cuboid $C$, with dimensions $X=\SI{100}{\meter}$, $Y=\SI{100}{\meter}$, $Z=\SI{20}{\meter}$. Each ground user was positioned in the center of a \SI{10}{\meter} $\times$ \SI{10}{\meter} subarea $a\in A$. The number of subareas was defined to consider the occupation of the base of cuboid $C$ (area $A$) equal to 5\%, 20\%, and 45\% of the total area available. Subarea $a\in A^s$ was characterized by a traffic demand $T^s$, where $T^{eMBB}$ was equal to \SI{20}{Mbit/s} and $T^{URLLC}$ was equal to \SI{4}{Mbit/s}, which correspond to, respectively, 25\% and 5\% of the data rate associated to the highest MCS index for \SI{20}{\mega\hertz} channel bandwidth, \SI{800}{\nano\second} GI, and single spatial stream wireless links. We considered $BER^{eMBB}$ equal to 10\textsuperscript{-5} and $BER^{URLLC}$ equal to 10\textsuperscript{-10}, which define the minimum SNR values required for transmitting a frame using any MCS index, taking into account the network configuration employed. Each networking scenario corresponds to a snapshot of the flying network at $t_k = k \cdot \Delta t, k \in \mathbb{N}_0$, where $\Delta t \gg \SI{1}{\second}$ is the flying network reconfiguration period. In a real-world deployment, each network reconfiguration implies resolving the optimization problem defined in \cref{eq:optimization-problem} for determining the up to date optimal solution. $\Delta t$ is a parameter that can be adjusted according to the dynamics of the networking scenario, in order to achieve a trade-off between the stability of the flying network, the conservatism of the performance requirements defined by the SLA, and the time required to determine the optimal solution for the placement and allocation of communications resources. For five networking scenarios composed of 45 subareas (the most complex networking scenarios considered), the average time spent by SLICER to solve the optimization problem defined in \cref{eq:optimization-problem} was $0.91 \pm 0.13$~\SI{}{\second} (95\% confidence interval), using an 11\textsuperscript{th} Generation Intel\textregistered{} Core\textsuperscript{TM} i5-1135G7 processor running at \SI{2.40}{\giga\hertz} and \SI{16}{\giga\byte} of RAM, which can be deployed on the Edge of the flying network. This computing time allows to meet the target $\Delta t \gg \SI{1}{\second}$ (e.g., $\Delta t$~=~\SI{30}{\second}). The fine-tuning of $\Delta t$ is left for future work.\looseness=-1

In the performance evaluation carried out, we considered two baseline approaches:
\begin{itemize}
    \item \textbf{\bm{$\left | S \right |$} FAPs (one FAP for each network slice \bm{$s \in S$})}, each placed in the geometric center of all subareas $a\in A^s$, at a random altitude of \SI{10}{\meter} or \SI{20}{\meter} (altitudes considered by SLICER), configured with up to \SI{160}{\mega\hertz} channel bandwidth.\looseness=-1
    \item \textbf{\bm{$\left | K \right |$} FAPs for each network slice \bm{$s \in S$}}, placed according to the state of the art k-means clustering algorithm~\cite{Likas2003}.
    First, the k-means clustering algorithm defines $\left | K \right |$ random positions as clusters' centroids. Then, it assigns each subarea $a\in A^s$ to the nearest cluster by calculating the distance to each centroid. After that, it determines the up to date centroid for each cluster by computing the average position among the assigned subareas. Each cluster's centroid defines the position where a FAP must be placed, at a random altitude of \SI{10}{\meter} or \SI{20}{\meter}. Each FAP must provide the cluster's subareas with the minimum channel bandwidth computed by SLICER for the same subareas, using up to \SI{160}{\mega\hertz} channel bandwidth. When any FAP does not have enough channel bandwidth available, $\left | K \right |$, initially set to 1, is successively increased by 1 and the k-means clustering algorithm is run again until all FAPs are able to provide the required bandwidth. The k-means clustering algorithm is run independently for each network slice $s \in S$.\looseness=-1
\end{itemize}

Two network slice types were considered:
\begin{itemize}
	\item \textbf{eMBB network slice} that aims at enabling the use of rich-media applications (e.g., video streaming) with average throughput equal to \SI{20}{Mbit/s} per ground user (subarea) and average delay up to \SI{5}{\milli\second}.\looseness=-1
	\item \textbf{URLLC network slice} that aims at enabling the use of mission-critical applications (e.g., communications for first-responders) with average throughput equal to \SI{4}{Mbit/s} per ground user (subarea) and average delay up to \SI{1}{\milli\second}.\looseness=-1
\end{itemize}

The two network slice types considered and the corresponding QoS levels were defined to evaluate and validate SLICER under network requirements imposed by representative applications. Yet, SLICER is valid for any number and type of network slices, and QoS levels. UDP Poisson, a source traffic model widely used for evaluating the performance of wireless networks, was considered.\looseness=-1

\subsection{Performance Metrics}

The evaluation presented herein takes into account three performance metrics:

\begin{itemize}
    \item \textbf{Throughput:} the number of bits received per second by the FAPs.
    \item \textbf{Packet Delivery Ratio (PDR):} the ratio between the number of packets received by the FAPs and the number of packets generated by the ground users.\looseness=-1
    \item \textbf{Delay:} the time taken by the packets to reach the sink application at the FAPs, considering as reference the time instant they were generated by the source application at the ground users.\looseness=-1
\end{itemize}

The performance metrics consist of average values for each second of all the simulation runs. They are represented by means of the Cumulative Distribution Function (CDF) for the delay and by the complementary CDF (CCDF) for the throughput and PDR. The CDF $F(x)$ represents the percentage of samples for which the delay is lower than or equal to $x$, while the CCDF $F'(x)$ represents the percentage of samples for which the throughput or PDR is higher than $x$.\looseness=-1

\begin{figure*}
	\centering
	\subfloat[Throughput CCDF.]{
	\includegraphics[width=0.30\linewidth]{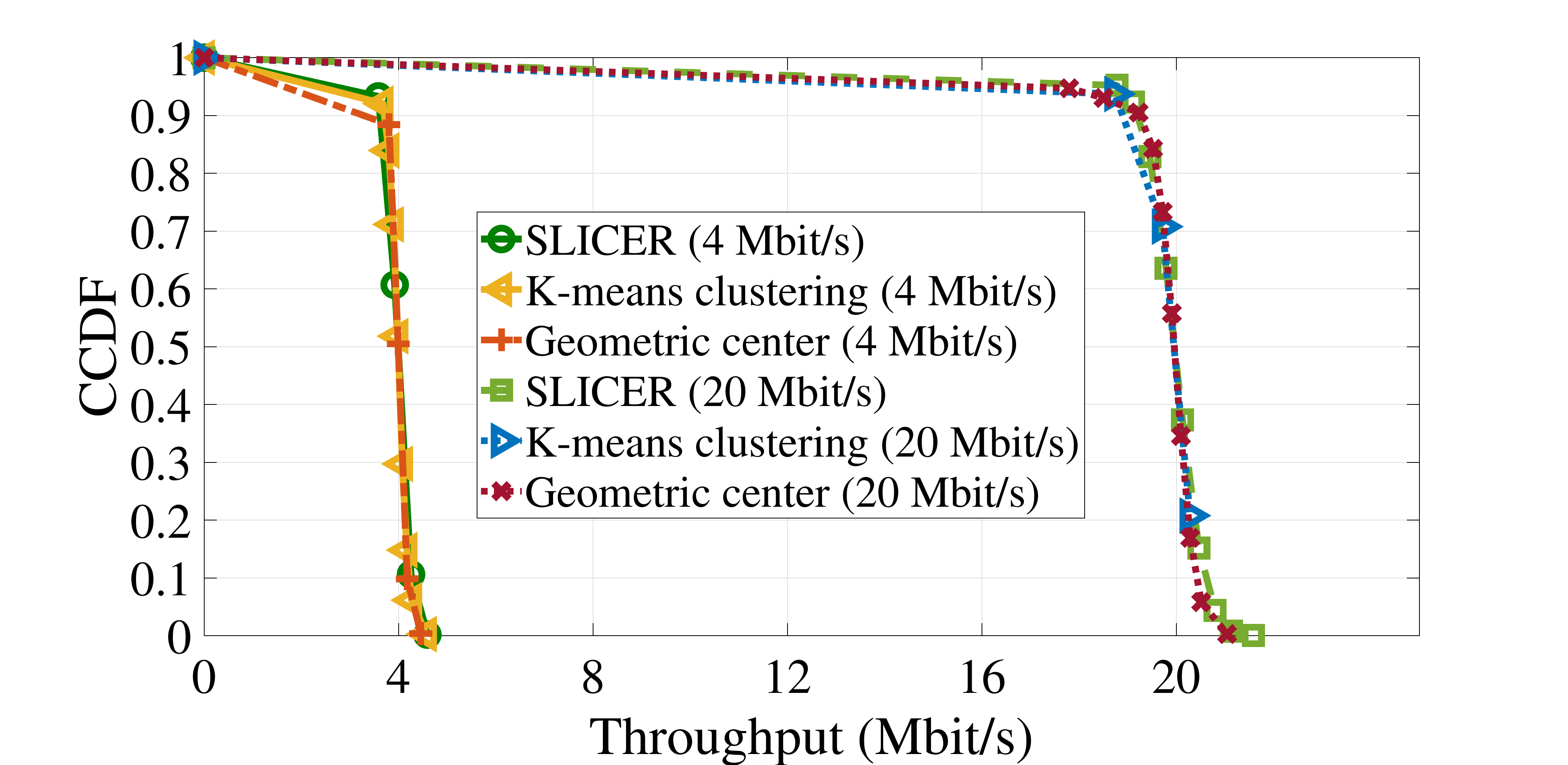}
	\label{fig:throughput-5subareas}}
	\subfloat[Packet Delivery Ratio (PDR) CCDF.]{
		\includegraphics[width=0.30\linewidth]{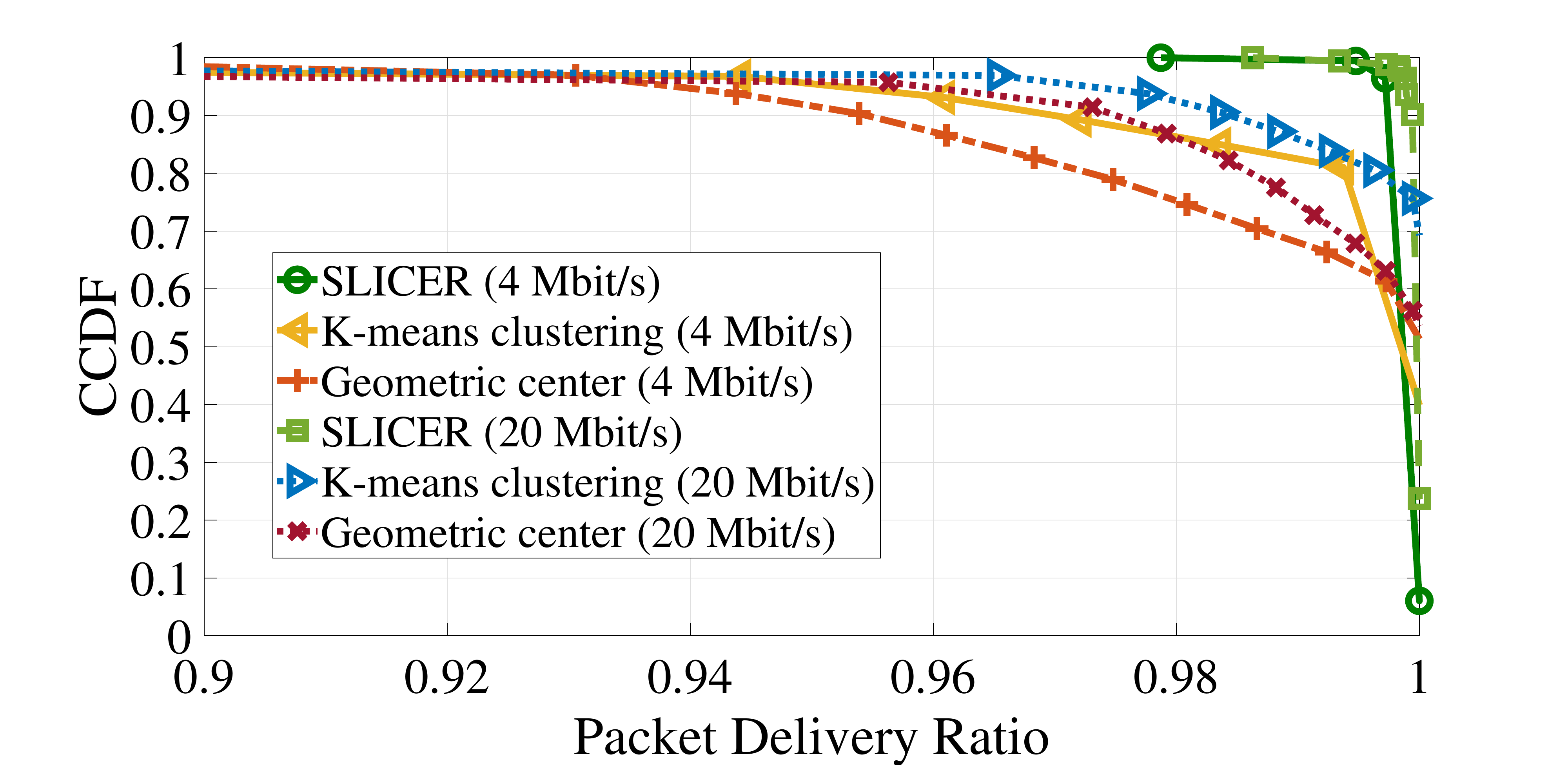}
		\label{fig:pdr-5subareas}}
	\subfloat[Delay CDF.]{
		\includegraphics[width=0.30\linewidth]{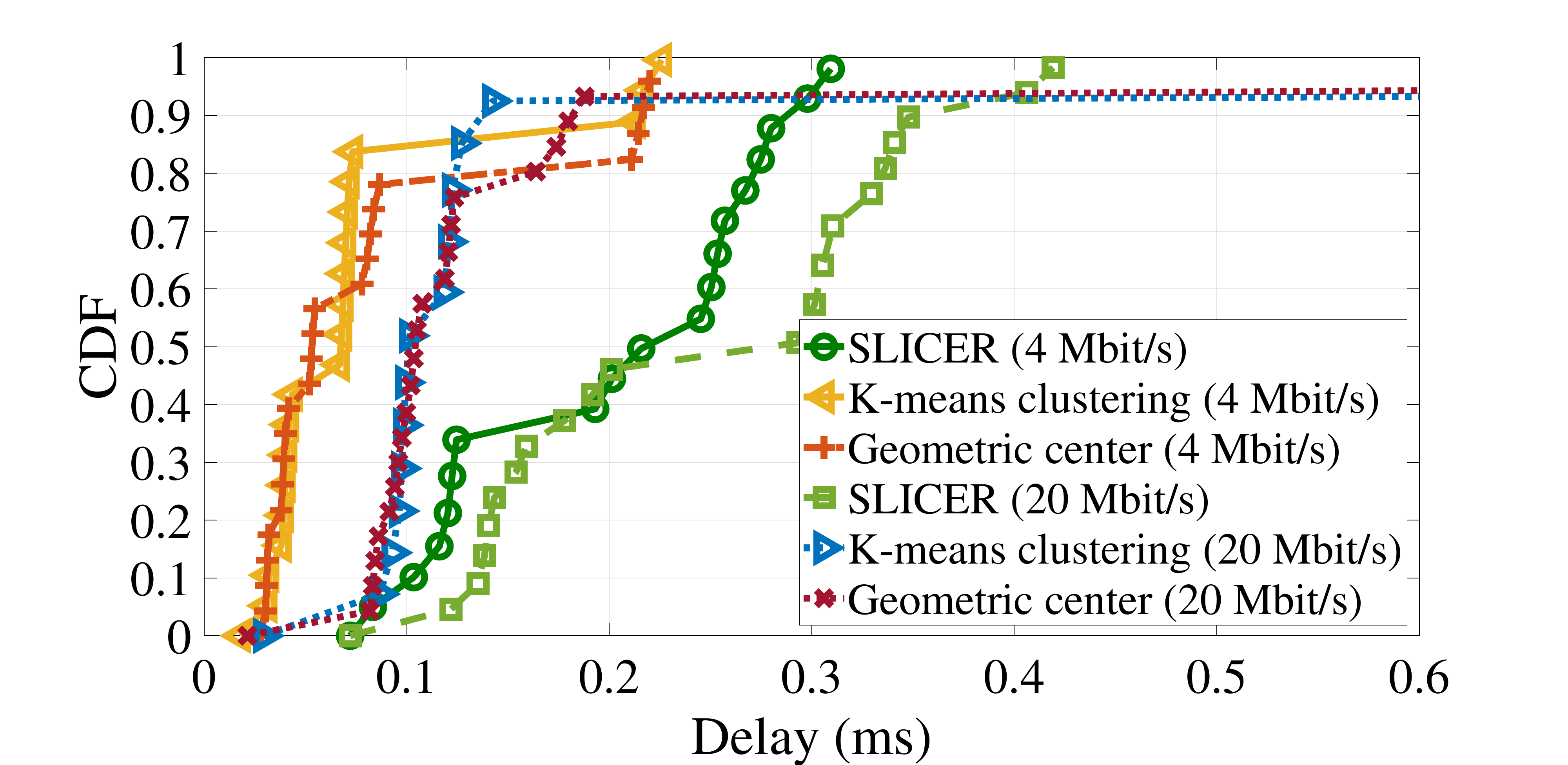}
		\label{fig:delay-5subareas}}
	\hfill
    \caption{Average performance results considering five networking scenarios composed of \textbf{5 subareas} randomly associated to URLLC and eMBB network slices.}
    \label{fig:performance-results-5subareas}
\end{figure*}

\begin{figure*} 
	\centering
	\subfloat[Throughput CCDF.]{
	\includegraphics[width=0.30\linewidth]{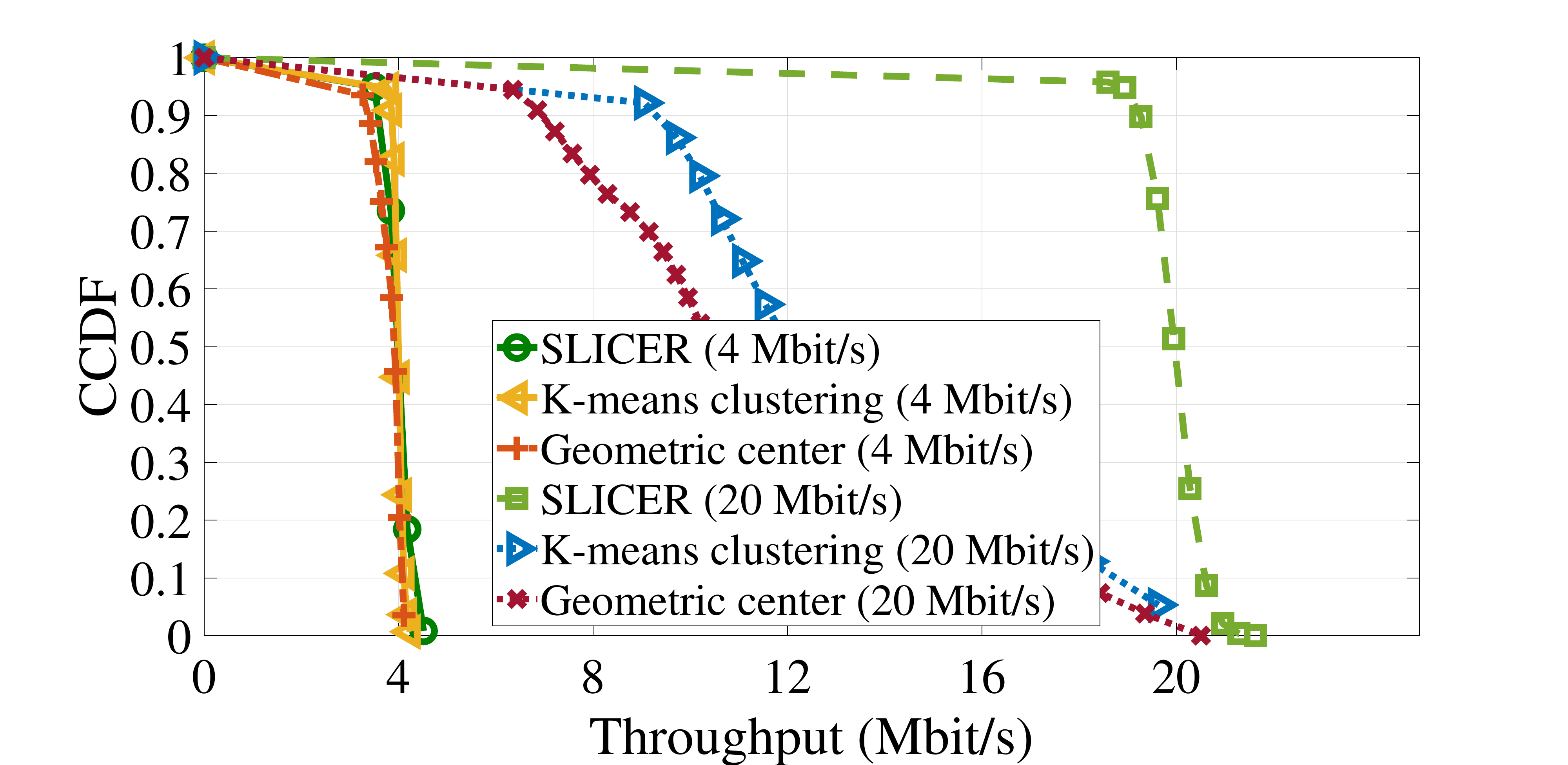}
	\label{fig:throughput-20subareas}}
	\subfloat[Packet Delivery Ratio (PDR) CCDF.]{
		\includegraphics[width=0.30\linewidth]{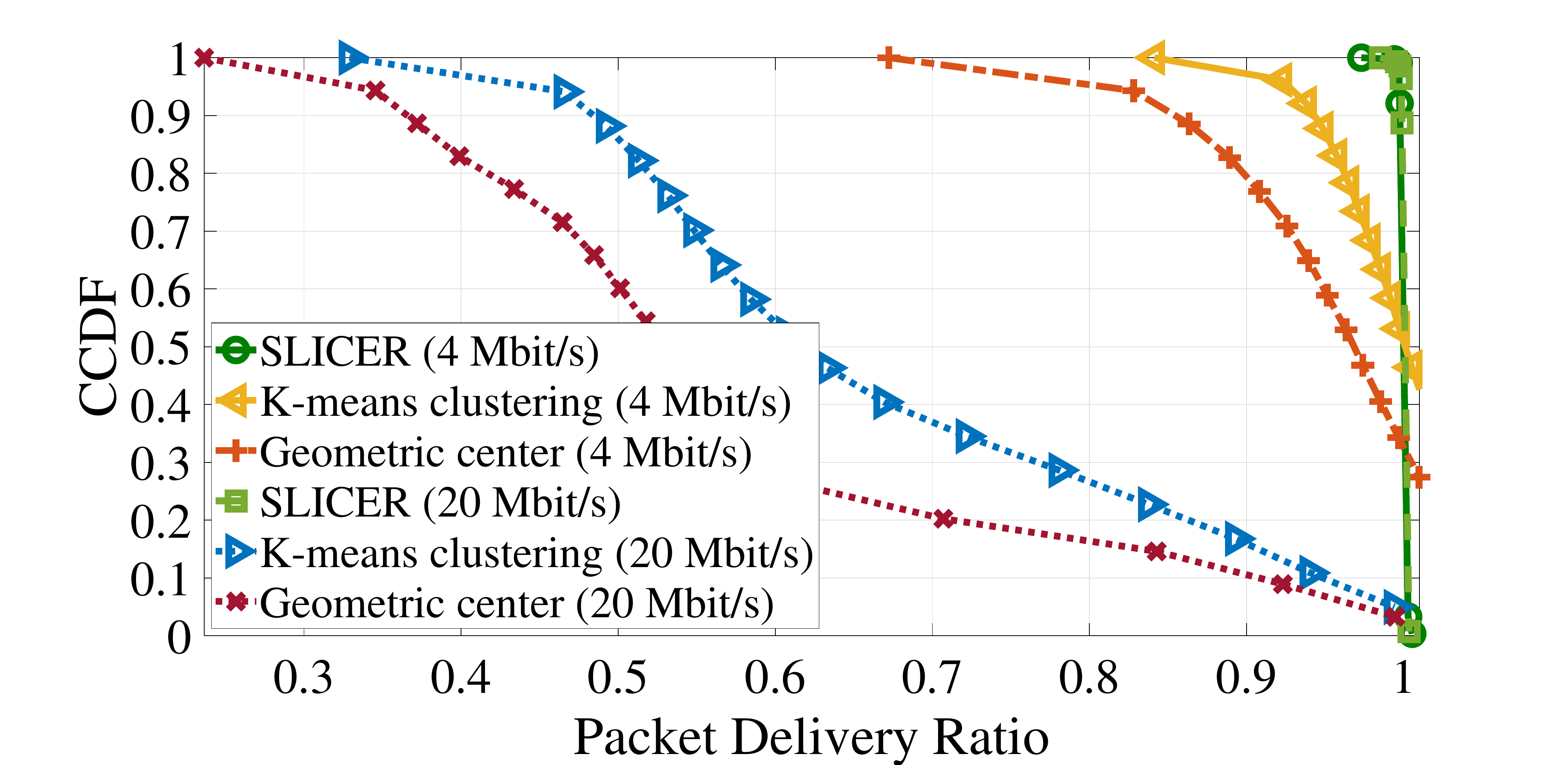}
		\label{fig:pdr-20subareas}}
	\subfloat[Delay CDF.]{
		\includegraphics[width=0.30\linewidth]{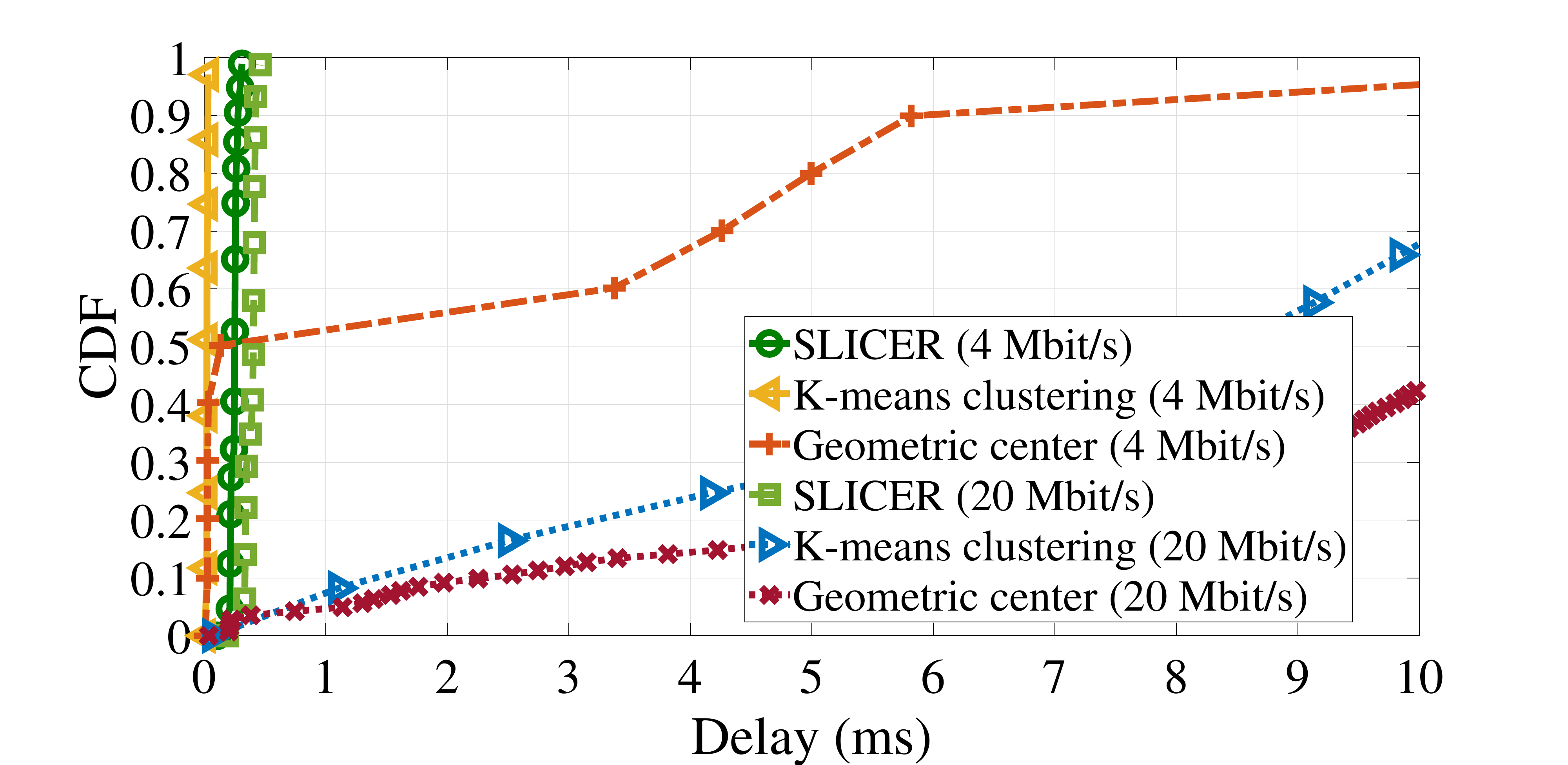}
		\label{fig:delay-20subareas}}
	\hfill
    \caption{Average performance results considering five networking scenarios composed of \textbf{20 subareas} randomly associated to URLLC and eMBB network slices.}
    \label{fig:performance-results-20subareas}
\end{figure*}

\begin{figure*}
	\centering
	\subfloat[Throughput CCDF.]{
	\includegraphics[width=0.30\linewidth]{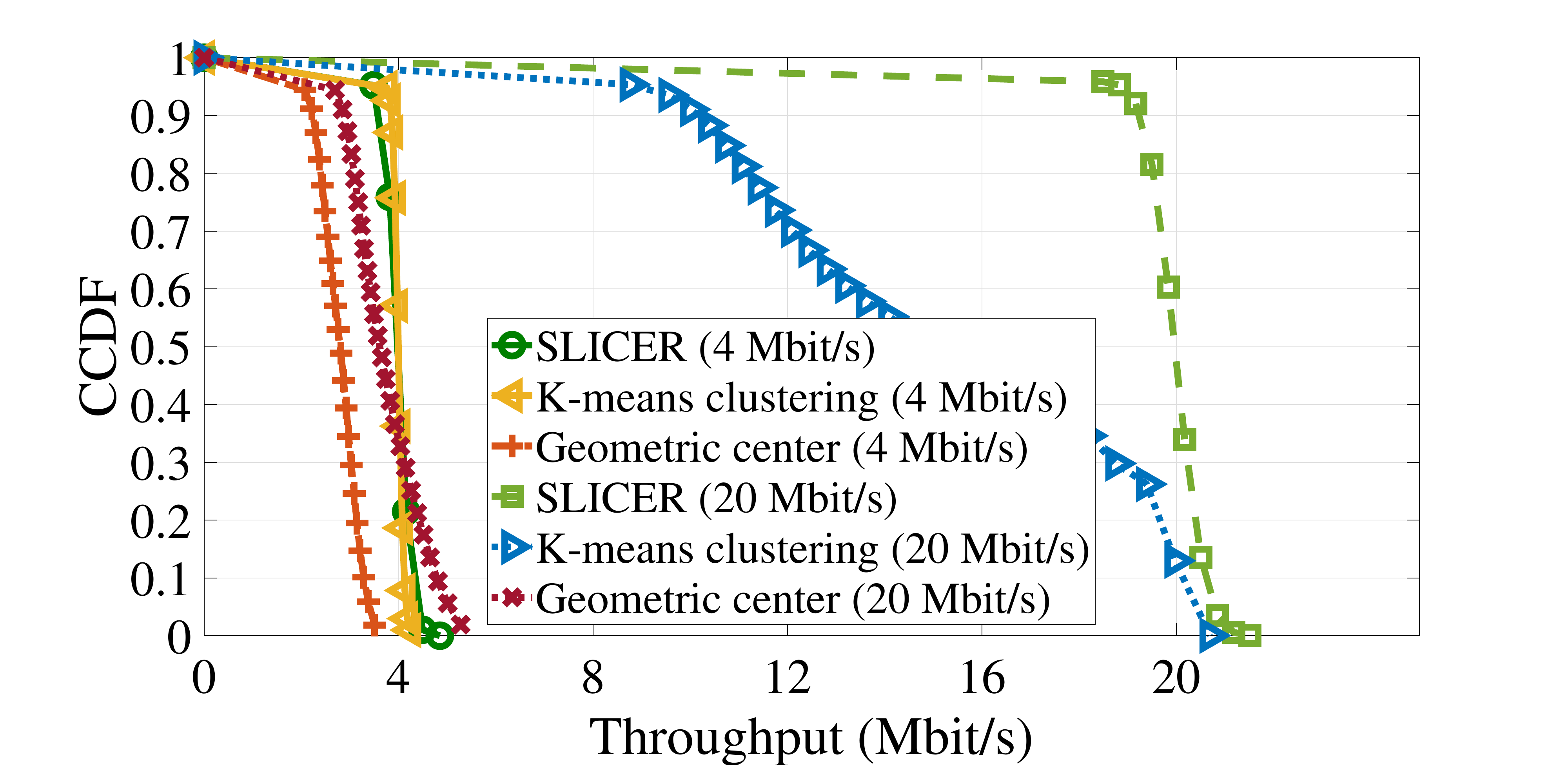}
	\label{fig:throughput-45subareas}}
	\subfloat[Packet Delivery Ratio (PDR) CCDF.]{
		\includegraphics[width=0.30\linewidth]{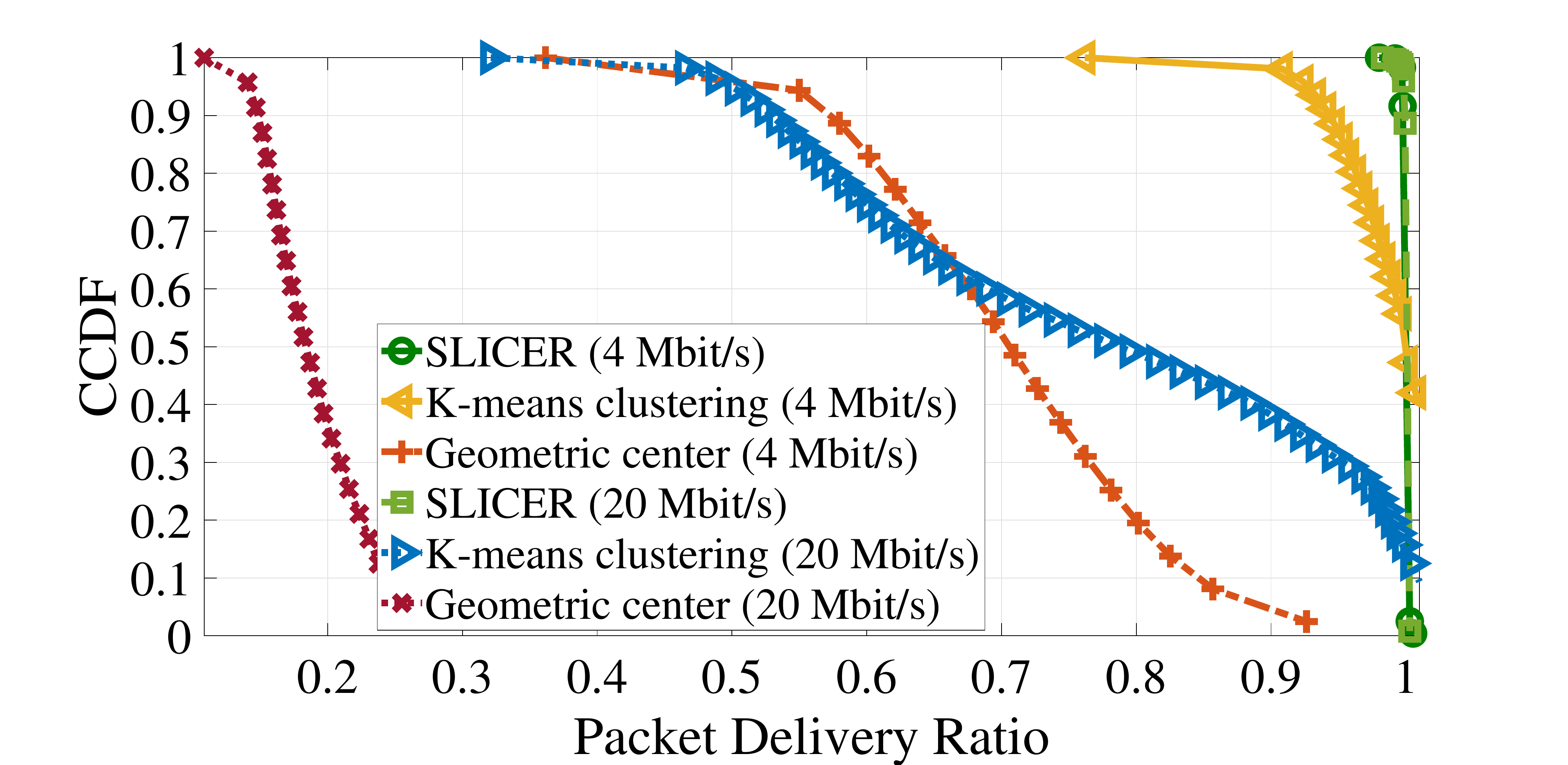}
		\label{fig:pdr-45subareas}}
	\subfloat[Delay CDF.]{
		\includegraphics[width=0.30\linewidth]{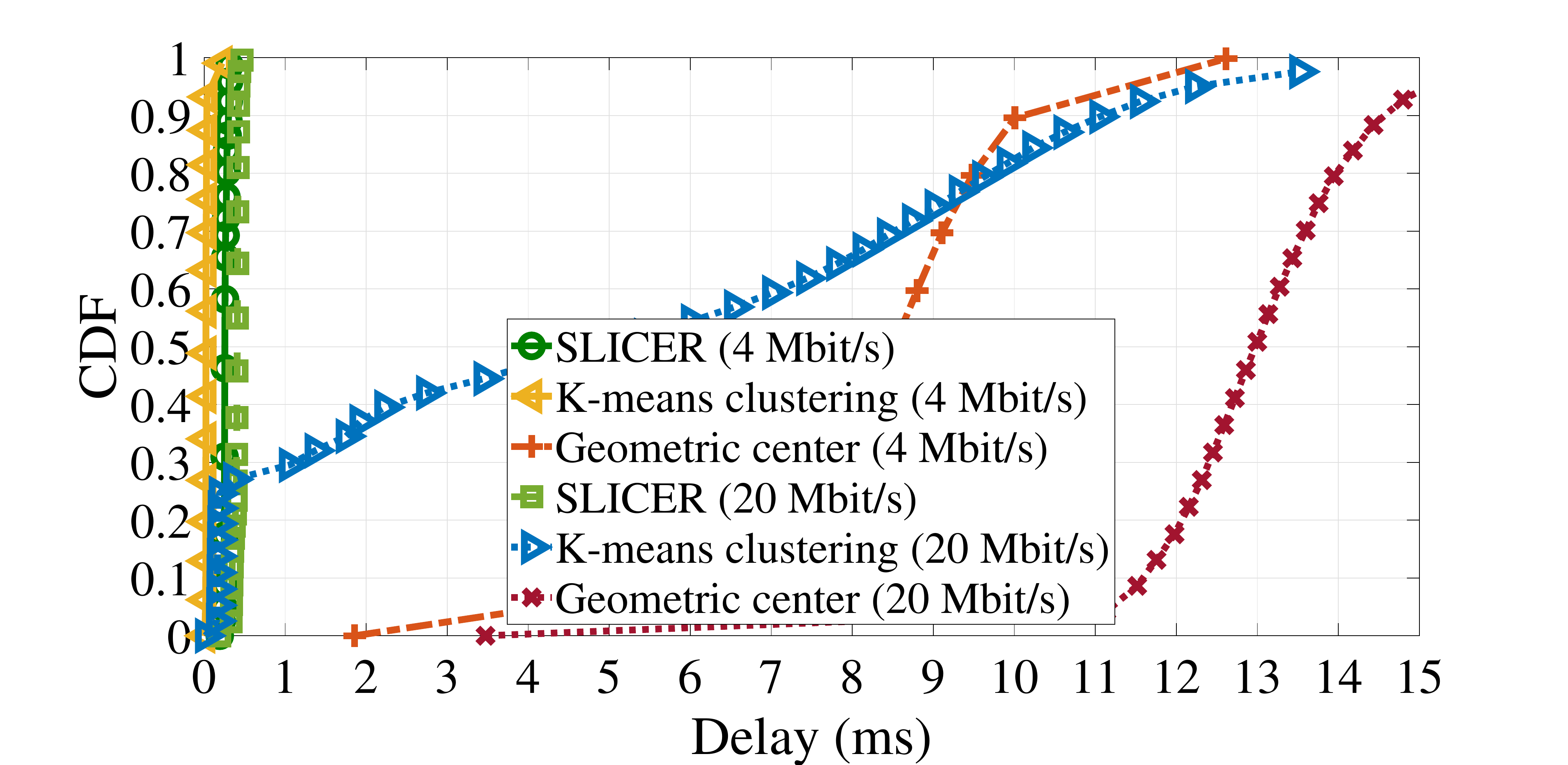}
		\label{fig:delay-45subareas}}
	\hfill
    \caption{Average performance results considering five networking scenarios composed of \textbf{45 subareas} randomly associated to URLLC and eMBB network slices.}
    \label{fig:performance-results-45subareas}
\end{figure*}

\begin{figure}
	\centering
	\subfloat[Average bandwidth.]{
	\includegraphics[width=0.47\linewidth]{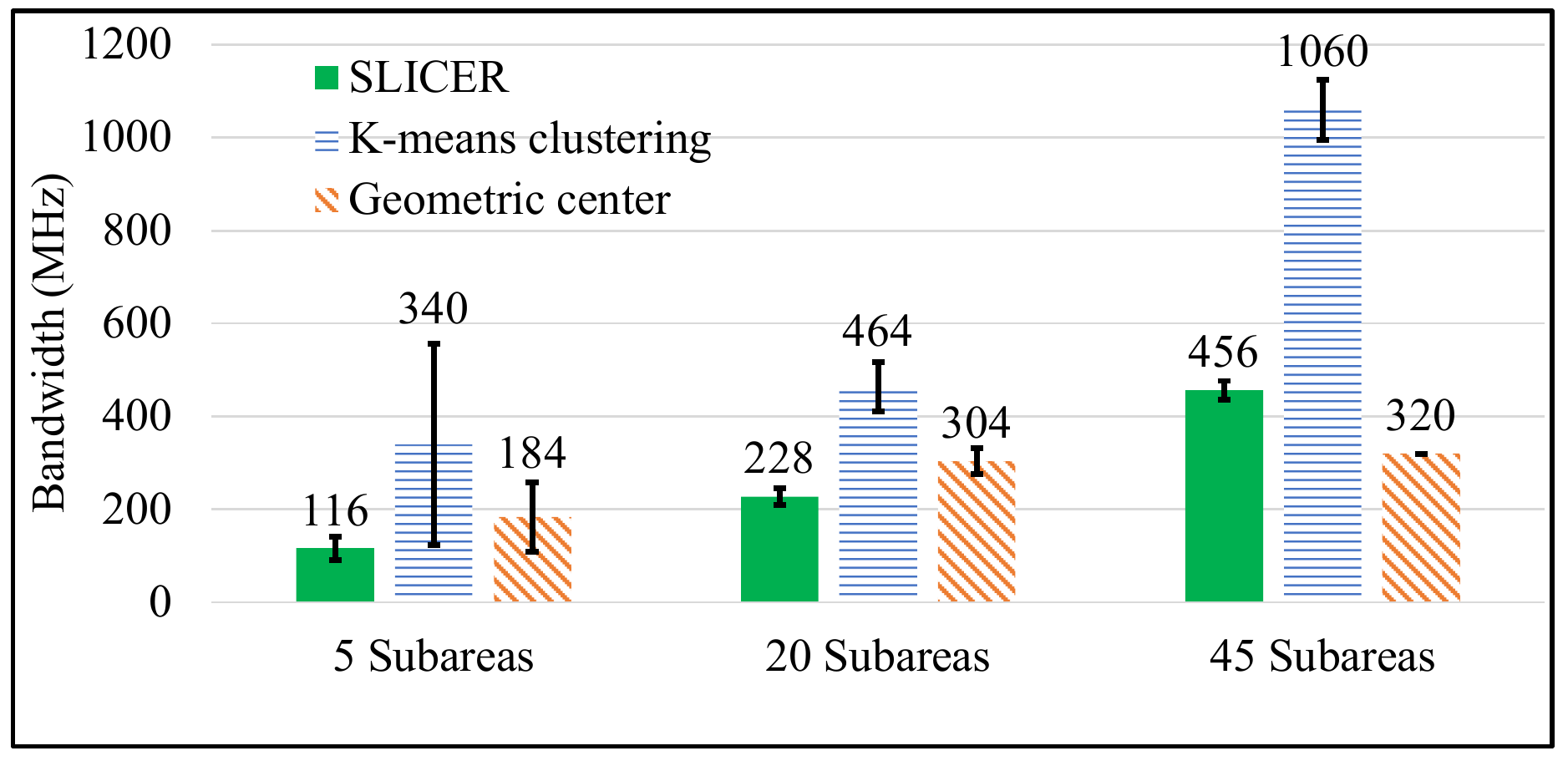}
	\label{fig:allocated-bw}}
	\hfill
	\subfloat[Average number of UAVs.]{
		\includegraphics[width=0.47\linewidth]{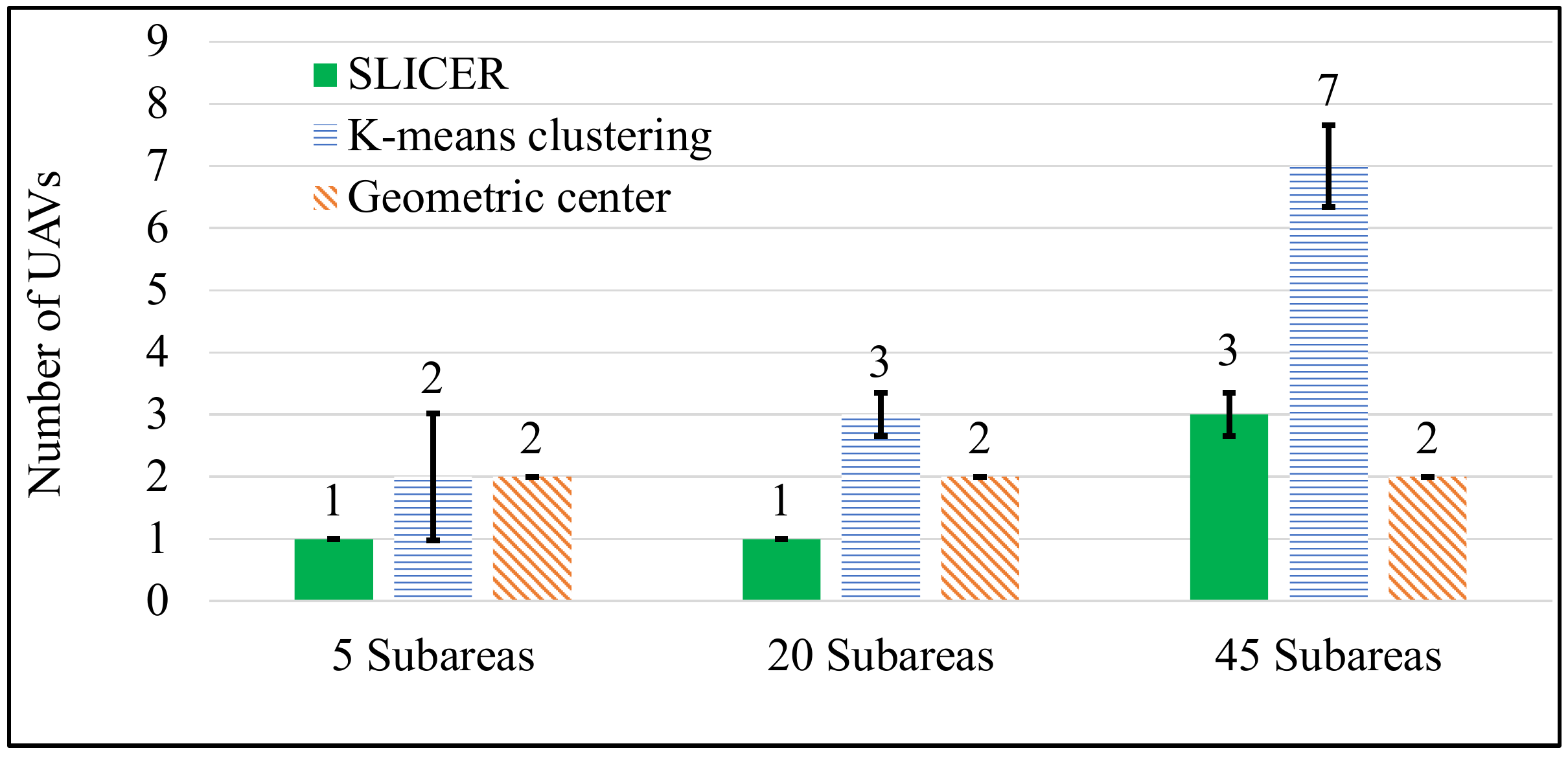}
	\label{fig:allocated-uavs}}
	\hfill
    \caption{Average bandwidth and number of UAVs used in networking scenarios composed of 5, 20, and 45 subareas randomly associated to URLLC and eMBB network slices, including 95\% confidence intervals.}
    \label{fig:bandwidth-and-uavs-used}
\end{figure}

\subsection{Simulation Results}
The simulation results obtained are presented in \cref{fig:performance-results-5subareas}, \cref{fig:performance-results-20subareas}, and \cref{fig:performance-results-45subareas}. They show that SLICER (cf. green circle and square markers) allows to meet the target QoS levels associated with the eMBB (\SI{20}{Mbit/s}) and URLLC (\SI{4}{Mbit/s}) network slices. For 5 subareas (cf.~\cref{fig:performance-results-5subareas}), all the solutions allow to meet the target QoS levels. This is due to the fact that, for this undemanding networking scenario, the k-means clustering and geometric center approaches offer over-provisioned communications resources, enabling PDR higher than 95\% and delay lower than \SI{0.2}{\milli \second}, when considering the 90\textsuperscript{th} percentile. They even outperform the delay achieved by SLICER. Nevertheless, it must be noted that SLICER does not violate the target QoS levels, as it was intended. On the other hand, the network performance achieved when using k-means clustering and geometric center is highly degraded when the number of subareas increases, which is clearly shown for eMBB (\SI{20}{Mbit/s}), considering 20 and 45 subareas (cf.~\cref{fig:performance-results-20subareas} and~\cref{fig:performance-results-45subareas}, respectively). K-means clustering is the solution that enables the network performance closer to SLICER, especially when the number of subareas increases.\looseness=-1

Regarding the resource usage in terms of average bandwidth and number of UAVs used (cf.~\cref{fig:bandwidth-and-uavs-used}), the results show SLICER uses only one UAV and requires the least amount of bandwidth for 5 and 20 subareas, when compared with the counterpart solutions. For 45 subareas, the average number of UAVs used by SLICER is three. Although the geometric center approach uses two UAVs, it leads to network performance degradation when the number of subareas increases, since assigning \SI{160}{\mega\hertz} channel bandwidth to each FAP, which is the maximum bandwidth available in current IEEE 802.11 standards, is not sufficient to meet the target QoS levels. On the other hand, the k-means clustering algorithm, which aims at providing to each subarea $a \in A^s$ the same minimum amount of bandwidth as SLICER, conduces in practice to a higher number of UAVs and amount of bandwidth used, since each FAP uses a single wireless channel. In addition, as the k-means clustering algorithm defines the clusters' centroids based on distance, computing the mean position among all subareas belonging to each cluster, it is not QoS-aware. From the communications point of view, it maximizes the SNR offered to all subareas belonging to the same cluster, but does not guarantee target SNR values. The problem is exacerbated when the wireless channel used by each FAP is not assigned to subareas belonging to different clusters, as SLICER does, which leads to wasted bandwidth in underused wireless channels.\looseness=-1

Despite the reduced amount of resources used, SLICER allows for better network performance. This is achieved by ensuring the optimized placement of the FAPs and the allocation of multiple channels with lower bandwidth (multiples of \SI{20}{\mega\hertz}), whereas the geometric center approach assigns to all subareas of each network slice a single wireless channel with up to \SI{160}{\mega\hertz} bandwidth, which may not be sufficient. On the other hand, the k-means clustering algorithm is the approach that uses the largest amount of resources. 
For this reason, it provides increased network performance than the geometric center approach. However, since the placement of the communications resources performed by the k-means clustering algorithm is not QoS-aware, the network performance achieved is worse when compared with SLICER.\looseness=-1

Although SLICER was formulated and validated in this paper for a flying access network only, it can also be employed in a backhaul network composed of UAV relays and gateways that forward the traffic between the FAPs and the Internet~\cite{Coelho2019}.\looseness=-1

\section{Conclusions\label{sec:Conclusions}}
We proposed SLICER, an algorithm enabling the on-demand placement and allocation of communications resources in slicing-aware flying networks. SLICER allows the computation of the minimum number of UAVs, their 3D positions, and the amount of communications resources to be provided in different geographical areas where network slices with target QoS levels must be made available. The flying network performance when using SLICER was evaluated by means of ns-3 simulations, considering multiple random networking scenarios. The obtained results show SLICER allows to meet the target QoS levels imposed by the network slices, while using the minimum amount of communications resources. 
As future work, we aim at developing a slicing-aware flying network prototype and evaluate the performance of SLICER in real-world networking scenarios. Moreover, we plan to evaluate the performance of SLICER using different optimization solvers.\looseness=-1

\section*{Acknowledgments}
This work is financed by the ERDF -- European Regional Development Fund through the Operational Programme for Competitiveness and Internationalisation -- COMPETE 2020 Programme and by National Funds through the Portuguese funding agency, FCT -- Fundação para a  Ciência e a Tecnologia under the PhD grant SFRH/BD/137255/2018.


\printbibliography

\end{document}